
%
%
%
%
%
%

\input harvmac

\def \sqr#1#2{\vcenter
{\hrule height .#2pt \hbox{\vrule width .#2pt
height #1pt\kern#1pt\vrule width .#2pt}\hrule height .#2pt}}
\def\minimatrix#1{\null\,\vcenter{\normalbaselines\mathsurround=0pt
    \ialign{\hfil$##$\hfil&&\enspace\hfil$##$\hfil\crcr
      \mathstrut\crcr\noalign{\kern-\baselineskip}
      #1\crcr\mathstrut\crcr\noalign{\kern-\baselineskip}}}\,}

\def \textorbsqr#1#2{\hbox{$\minimatrix{{#1}&\sqr{8}{8}\cr
                                  {}&{#2}\cr}$}}

\def \orbsqr#1#2{\hbox{$\minimatrix{{ }&{ }\cr
                                {#1}&\sqr{12}{12}\cr
                                {}&{#2}\cr}$}}

\def \MMorbsqr#1#2{\hbox{$\minimatrix{{ }&{ }\cr
                                {#1}&\sqr{12}{12}^{\ \natural}\cr
                                {}&{#2}\cr}$}}
\def \MMtextorbsqr#1#2{\hbox{$\minimatrix{{ }&{ }\cr
                                {#1}&\sqr{8}{8}^{\ \natural}\cr
                                {}&{#2}\cr}$}}

\def \subsect#1{\medskip\noindent {\bf #1}}

\def\pfu{partition function\ }
\def\pf{partition function}

\def\auto{automorphism\ }
\def\autos{automorphisms\ }
\def\mod{modular\ }

\def\orb{orbifold\ }
\def\op{operator\ }
\def\ops{operators\ }

\def\Tr#1#2{{\rm Tr}_{#1}(#2)}
\def\Torb#1{T_{#1}^{\rm orb}(\tau)}
\def\det#1{{\rm det}(#1)}

\def\MM{{\cal V}^{\natural}}
\def\a{{\overline a}}

\def\c{{\overline c}}
\def\r {{\overline r}}
\def\Tgt{T_g(\tau)}
\def\L{\Lambda}
\def\VL{{\cal V}^\L}
\def\Vr{{\cal V}_r}
\def\Va{{\cal V}_a}
\def\Vp {{\cal V}'}
\def\Vak{{\cal V}_{a^k}}
\def\Vg{{\cal V}_g}

\def\Vorb{{\cal V}_{\rm orb}^a}

\def\MMforb{{\cal V}_{\rm orb}^f}
\def\Morb{M_{\rm orb}^a}

\def\La{L_\a}

\def\Lahat{\hat\La}
\def\Pr{{\cal P}_r}
\def\Pa{{\cal P}_a}

\def \Tft {T_f(\tau)}
\def \Vf{{\cal V}_{f}}
\def \Vfk{{\cal V}_{f^k}}

\def\Co{{\rm Co}_1}
\def\G{\Gamma}
\def\Gg{\G_g}
\def\Ga{\G_a}
\def\GN{\G_0(N)}
\def\Gn{\G_0(n)}
\def\Gp{\G_0(p)}

\lref\FLM{
Frenkel, I.,  Lepowsky, J.  and Meurman, A.,
 Proc.Natl.Acad.Sci.USA {\bf 81} (1984) 3256;
J.Lepowsky et al. (eds.), Vertex operators in mathematics and physics,
(Springer Verlag, New York, 1985);
Vertex operator algebras and the monster,
(Academic Press, New York, 1988).}

\lref\Dixon{
Dixon, L.,   Harvey, J.A., Vafa, C.,  and Witten, E.,
Nucl.Phys. {\bf B261} (1985) 678;
Nucl.Phys. {\bf B274} (1986) 285.}

\lref\DGH{
Dixon, L. Ginsparg, P.  and Harvey, J.A.,
Comm.Math.Phys. {\bf 119} (1988) 285.}

\lref\CS{
Conway, J.H.  and Sloane,  N.J.A.,
Sphere packings, lattices and groups,
(Springer Verlag, New York, 1988).}

\lref\GO{
Goddard, P.  and Olive, D.,
J.Lepowsky et al. (eds.), Vertex operators in mathematics and physics,
(Springer Verlag, New York, 1985).}

\lref\Serre{
Serre, J-P.,
A course in arithmetic,
(Springer Verlag, New York, 1970).}

\lref\Goddard{
Goddard, P.,
Proceedings of the CIRM Luminy conference, 1988,
(World Scientific, Singapore, 1989).}

\lref\DVVV{
Dijkgraaf, R.,   Vafa, C., Verlinde,  E. and  Verlinde, H.,
Comm.Math.Phys. {\bf 123} (1989) 485.}

\lref\CHZtwo{
Corrigan, E.   and  Hollowood, T.J.,
Nucl.Phys. {\bf B304} (1988) 77.}

\lref\DGMZtwo{
Dolan, L.,  Goddard, P. and Montague, P.,
Nucl.Phys. {\bf B338} (1990) 529.}

\lref\DFMS{
Dixon, L.,  Friedan,  D., Martinec, E.  and Shenker,  S.,
Nucl.Phys. {\bf B282} (1987) 13.}

\lref\Vafa{
Vafa, C.,
Nucl.Phys. {\bf B273} (1986) 592.}

\lref\Ginsparg{
Ginsparg, P.,
Les Houches, Session XLIX, 1988, "Fields, strings and critical phenomena", ed.
E. Brezin and J. Zinn-Justin,
 Elsevier Science Publishers (1989).}

\lref\Griess{
Griess, R.,
Inv.Math. {\bf 68} (1982) 1.}

\lref\Thompson{
Thompson, J.G.,
Bull.London Math.Soc.{\bf 11} (1979) 347.}

\lref\Tuitetwo{
Tuite, M.P.,
DIAS-STP-90-30,  To appear in Commun.Math.Phys..}

\lref\DongMason{
Dong, C.  and Mason, G.,
U.C.Santa Cruz Preprint 1992.}

\lref\CN{
Conway, J.H. and  Norton, S.P.,
Bull.London.Math.Soc. {\bf 11} (1979) 308.}

\lref\Twist{
Lepowsky, J.,
Proc.Natl.Acad.Sci.USA {\bf 82} (1985) 8295;
Kac, V.  and Peterson, D.,
Proceedings of the Argonne symposium on anomolies, geometry, topology, 1985
(World Scientific, Singapore, 1985).;
Corrigan, E.   and  Hollowood, T.J.,
Nucl.Phys. {\bf B303} (1988) 135.}

\lref\Kondo{
Kondo, T.,
J.Math.Soc.Japan {\bf 37} (1985) 337.}

\lref\Borch{
Borcherds, R.,
Univ. Cambridge DPMMS preprint 1989.}

\lref\Wilson{
Wilson, R.,
J.Alg. {\bf 85} (1983) 144.}

\lref\DGMtri{
Dolan, L.,  Goddard, P.  and Montague, P.,
Phys.Lett. {\bf B236} (1990) 165.}

\lref\Tuiteone{
Tuite, M.P.,
Commun.Math.Phys. {\bf 146} (1992) 277.}

\lref\Ogg{
Ogg, A.,
Bull.Soc.Math.France {\bf 102} (1974) 449.}

\lref\Unique{
Tuite, M.P.,
DIAS preprint 1992, in preparation.}

\lref\Tuitethree{
Tuite, M.P.,
DIAS-STP-91-25  Aug 1991.}

\lref\Gunning{
Gunning, R.C.,
Lectures on modular forms,
(Princeton University Press, Princeton, 1962).}

\hfill DIAS-STP-92-29\break
\vskip 1 truecm
\centerline {\bf MONSTROUS MOONSHINE  AND  }
\centerline {\bf THE UNIQUENESS OF THE MOONSHINE MODULE
\footnote * {Talk presented at the Nato Advanced Research Workshop
on \lq Low dimensional topology and quantum field theory\rq,
Cambridge, 6-13 Sept 1992.}}
\vskip 1truecm
\centerline{Michael P. Tuite \footnote \dag{EMAIL: mphtuite@bodkin.ucg.ie}}
\centerline{Department of Mathematical Physics}
\centerline {University College}
\centerline {Galway, Ireland}
\smallskip
\centerline {and}
\smallskip
\centerline{Dublin Institute for Advanced Studies}
\centerline{10 Burlington Road}
\centerline{Dublin 4, Ireland}
\vskip 1.5 truecm

\centerline {ABSTRACT}
\medskip
In this talk we consider the relationship between the conjectured uniqueness
of the Moonshine module $\MM$ of Frenkel, Lepowsky and Meurman
 and Monstrous Moonshine, the genus zero property for
Thompson series discovered by Conway and Norton. We discuss some
evidence to support the uniqueness of $\MM$  by considering possible
alternative orbifold constructions of $\MM$ from a Leech lattice compactified
string. Within these constructions we find a new relationship between the
centralisers of the Monster
group and the Conway group generalising an observation made by Conway and
Norton. We also relate the uniqueness of $\MM$ to Monstrous Moonshine
and argue that given this uniqueness, then the genus zero properties hold
if and only if orbifolding $\MM$ with respect to a monster element reproduces
$\MM$ or the Leech theory.

\vfill
\eject
\pageno =1

\subsect {The Moonshine Module.}
The Moonshine module \FLM\ of Frenkel, Lepowsky and Meurman (FLM)
is the first example of an orbifold CFT \Dixon\ and  is
constructed from a string compactified to $R^{24}/\L$
where $\L$ is the Leech lattice, the unique 24 dimensional even self-dual
lattice without roots i.e. $\lambda^2\not = 2 $ cf.
\CS. The orbifolding is then based on the $Z_2$  reflection \auto of $\Lambda$.

Let $\VL$ denote the set of vertex operators  $\{\phi (z)\}$ for the Leech
lattice CFT which forms a closed meromorphic  \op product algebra (OPA) with
central charge 24 \refs{\FLM, \Goddard}
$$
\eqalign{
\phi_i(z)\phi_j(w)
\sim \sum_k C_{ijk}(z-w)^{h_k-h_i-h_j}\phi_k(w)+...}
\eqno(1)
$$
We will represent such an OPA schematically by $\phi \phi \sim \phi$.
The 1-loop  \pfu  $Z(\tau)=\Tr{\VL}{q^{L_0}}$ is a modular invariant
and meromorphic function of $\tau$ with a unique simple pole at
$q= e^{2\pi i \tau}= 0$ and is given by the unique (up to an additive constant)
modular invariant function $J(\tau)$
$$
\eqalign{Z(\tau)&=J(\tau)+24\cr
J(\tau)&={E_2^3\over{\eta^{24}}}-744
={1\over q}+0+196884q+...\cr}
\eqno(2)
$$
The constant 24 reflects the existence of
24 massless (conformal weight 1) \ops in this theory.
$\eta(\tau)=q^{1/24}\prod_n (1-q^n)$ and  $E_2(\tau)$ is the
Eisenstein modular form of weight 4 \Serre.

The FLM Moonshine module \FLM\ is an \orb CFT based on the
 $Z_2$ lattice reflection
\auto $\r:\lambda\rightarrow -\lambda$ for $\lambda\in\Lambda$. $\r$ lifts
to a family of $Z_2$ \autos of $\VL$ preserving (1)
from which family one \auto $r$ is chosen. Defining the projection
$\Pr=(1+r)/2$, the set of \ops $\Pr \VL$ then
also form a closed meromorphic OPA. However, the corresponding \pfu
$\Tr{\Pr\VL}{q^{L_0}}=
{1\over 2}(\textorbsqr{1}{1}+\textorbsqr{r}{1})$ is not modular invariant,
 employing standard notation for the world-sheet torus boundary conditions
e.g. \refs{\Ginsparg}. Thus, under a \mod transformation $\tau\rightarrow
-1/\tau$, $\textorbsqr{r}{1}={1/\eta_\r(\tau)}\rightarrow
\textorbsqr{1}{r}=2^{12}\eta_\r(\tau/2)=2^{12}q^{1/2}+...$
where $\eta_\r(\tau)=[\eta(2\tau)/\eta(\tau)]^{24}$. Therefore the introduction
of a \lq twisted\rq\ sector with vacuum energy $1/2$ and degeneracy $2^{12}$ is
necessary  to form a modular invariant theory \refs{\FLM,\Dixon}.
The states of this sector  are constructed from  twisted vertex \ops
$\Vr=\{\psi(z)\}$ acting on the untwisted vacuum. Thus $\VL$ is enlarged
by $\Vr$ to $\Vp=\VL\oplus\Vr$ which forms a closed non-meromorphic OPA
\refs{\FLM,\DFMS,\CHZtwo,\DGMZtwo} where (schematically)
$$
\eqalign{\phi\phi\sim \phi, \qquad
 \phi \psi\sim  \psi, \qquad  \psi \psi \sim  \phi}
\eqno{(3)}
$$
$\r$ can also be lifted to an \auto $r$ of (3) where the \ops of $\Pr\Vr$
have integral conformal weight. Then  $\MM=\Pr\Vp$ forms a closed
meromorphic OPA, the FLM Moonshine module \FLM. The $r$ projection ensures
the absence of untwisted massless \ops whereas the twisted sector \ops
are all massive since the twisted vacuum energy is $1/2$.
Thus the orbifold \pfu is
$$
\eqalign{
\Tr{\MM}{q^{L_0}}=\orbsqr{\Pr}{1}+\orbsqr{\Pr}{r}=J(\tau)}
\eqno (4)
$$

The absence of massless \ops in $\MM$ sets the Moonshine module
apart from other CFTs. Usually such \ops are present and form a closed
Kac-Moody algebra. However, the 196884 conformal weight 2 operators
in $\MM$, including the energy-momentum tensor $T(z)$ can be used to define
a closed non-associative commutative algebra. FLM demonstrated \FLM\ that
this algebra is an affine version of the 196883 dimensional Griess algebra
\Griess\ together with $T(z)$. The \auto group of the Griess algebra
is the Monster $M$. FLM showed that $M$ is the \auto group for the OPA
of $\MM$ where the \ops of $\MM$ of a given
conformal weight form a (reducible) representation
of $M$. This demonstrates an observation of McKay and Thompson
\Thompson\ that the coefficients of $J(\tau)$ are positive sums of dimensions
of
 irreducible representations of $M$ e.g. the coefficient of $q$ is
$196884=1+196883$, the sum of the trivial and adjoint representation.

We may  identify an involution $i\in M$, defined like a \lq fermion number\rq,
under which all untwisted (twisted) \ops have eigenvalue $+1( -1)$ where $i$
also respects (3). The centraliser of $i$
can be found  \FLM\ to give $C(i\vert M)=\{ g\in M\vert ig=gi\}=
2^{1+24}_{+}.\Co$ where $\Co$ is the Conway simple
group (the \auto group $\rm Co_0$ of $\L$ modulo the reflection \auto $\r$),
$2^{1+24}_{+}$ is an extra-special group
and $A.B$ denotes a group with normal subgroup $A$ with $B=A.B/A$. This result
 is an essential part of the FLM construction since $M$ is generated by
$2^{1+24}_{+}.\Co$ and a second involution  $\sigma$ \Griess. FLM constructed
$\sigma$, which mixes the untwisted and twisted sectors, from a hidden triality
symmetry  \refs{\FLM,\DGMtri}
and hence showed that the \auto group of $\MM$ is $M$.

The \autos $i$ and $r$ can be said to be \lq dual\rq\ to each other in the
sense that they are
both \autos of $\Vp$ and that the subsets invariant under
$i$ and $r$, $\VL$ and $\MM$ repectively, form meromorphic OPAs.
In addition, we may \lq reorbifold\rq\ $\MM$ with respect to $i$ to reproduce
$\VL$. Thus
$$
\eqalign{
\matrix{
 \Vp  \cr
{\buildrel {\cal P}_{i}\ \  \over\swarrow} \qquad {\buildrel \ \
\Pr\over\searrow}  \cr
 \VL\quad
         \matrix{{\buildrel r\over \longrightarrow}  \cr
                {\buildrel i\over \longleftarrow}}
\quad\MM\cr}
}
\eqno (5)
$$
where the horizontal arrows denote an orbifolding and the diagonal arrows a
projection \Unique.

\subsect {Monstrous Moonshine.}
The \ops of $\MM$ of a given conformal weight form reducible representations of
$M$. The Thompson series $\Tgt$ for $g\in M$  is defined by the trace
$$
\eqalign{
\Tgt=\Tr{\MM}{gq^{L_0}}={1\over q}+0+[1+\chi(g)]q+...
}
\eqno (6)
$$
which depends only on the conjugacy class of $g$ where $\chi(g)$ is the
character in the 196883 dimensional irreducible representation.
Thus for $i$ defined above, it is
easy to show $T_i(\tau)=[\eta_\r(\tau)]^{-1}+24$.

The Thompson series for the identity element is $J(\tau)$ which is unique
(up to a constant) for the  following reasons. Let ${\cal F}=H/\G$ be the
fundamental region where $\G={\rm SL}(2,Z)$ is the full modular group
and $H$ is the upper half complex plane. Adding the point at infinity,
the compactification $\overline{\cal F}$ is isomorphic to the Riemann sphere of
genus zero where the function $J(\tau)$ realises this isomorphism.
Such a function is called a {\it hauptmodul for the genus zero modular group
$\G$}. A modular invariant meromorphic function is a hauptmodul if and
only if it
possesses a unique simple pole. Once the location of this pole is specified,
this function is itself unique up to a constant cf. \refs{\Serre, \Tuiteone}.

Based on \lq experimental\rq\ evidence, Conway and Norton \CN\ conjectured
that each $\Tgt$ is a hauptmodul for a genus zero
modular group $\Gg$.  This has recently been rigorously demonstrated by
Borcherds although the origin of the genus zero property remains obscure
\Borch.
In general, for $g$ of order $n$,
$\Tgt$ is found to be invariant up to phases of order (at most) $h$ under
$\G_0(n)=\{(\matrix{a& b\cr
              nc & d\cr})\vert {\rm det }=1\}$ where $h\vert n$ and $h\vert
24$.
$\Tgt$ is fixed by $\Gg$ with $\GN\subseteq \Gg
\subseteq {\cal N}(N)=\{\rho\in {\rm SL}(2,R)\vert \rho \GN =\GN \rho \}$,
the normaliser of $\GN$ in ${\rm SL}(2,R)$ where $N=nh$.
Furthermore, $\Gg$ is a genus zero modular group and
$\Tgt$ is the corresponding hauptmodul with a simple pole at $q=0$.
Consider the elements of prime order $n=p$. Apart from one class
of order 3 with $h=3$, we have $h=1$ in each case.
Thus either $\Gg=\Gp$ or $\Gp+$,  generated by $\Gp$ and
the Fricke involution $W_p:\tau\rightarrow -1/p\tau $ with  $W_p^2=1$,
the only non-trivial element of ${\cal N}(p)$.  $\Gp$ is of genus zero when
$(p-1)\vert 24$  ($p=2,3,5,7,13$) where the hauptmodul is
$[\eta(\tau)/\eta(p\tau)]^{2d}+2d$ with $2d=24/(p-1)$.
There is a class of $M$ denoted
by $p-$ for each such prime with this Thompson series e.g. the involution
$i$ belongs to the class $2-$.
$\Gp+$  is of genus zero for $2\leq p \leq 31$ or  $p=41,47,59,71$,
which constitute all the prime divisors of the order of $M$ \Ogg.
Similarly, there is a class of $M$, denoted by $p+$, for each such prime with
Thompson series fixed by $\Gp+$.

It is natural to interprete the Thompson series $\Tgt$ as a contribution to the
\pfu for a further orbifolding of $\MM$ with respect to $g$
\refs{\DGH,\Tuiteone}.
In particular, we expect that under $\tau\rightarrow-1/\tau$, $T_g(\tau)$
transforms to the partition function for a $g$ twisted sector $\Vg$ as follows:
$$
\eqalign{
T_g(\tau)=\MMorbsqr{g}{1}\rightarrow \MMorbsqr{1}{g}
	=N_gq^{E_0^g}+...
}
\eqno (7)
$$
where $\natural$ denotes a trace contribution to the orbifolding
of $\MM$ and $\Vg$ has vacuum energy $E_0^g$ and
degeneracy $N_g$. For many classes of $M$, the method of construction of
$\Vg$ is not known. However, for certain elements discussed below and
some others, a construction can be given \refs{\Tuiteone,\Unique}.

Consider now this orbifold picture of $\Tgt$ for the
prime classes $p+$ and $p-$, although
 the analysis given can be generalised to all classes
\refs{\Tuiteone,\Tuitetwo,\Unique}. Under a modular
transformation $\gamma :\tau\rightarrow {a\tau+b\over c\tau+d}$ we find
$\MMtextorbsqr{g}{1}\rightarrow \MMtextorbsqr{g^{-d}}{g^{c}}$
assuming that no extra global phase occurs \Vafa \ (such a phase corresponds
to $h\not =1$  in the original Moonshine conjectures \refs{\Tuiteone,\Unique}).
For $\gamma\in \Gp$ with $c= 0\ {\rm mod}\ p$ we find
$\gamma :\Tgt\rightarrow T_{g^{-d}}(\tau)=\Tgt$ since $d$ and $p$ are
relatively prime and $\Tgt$ is  $\Gp$ invariant.

The genus zero property can be also understood as follows.
$\Tgt$ always has a simple pole at $q=0\ (\tau=\infty)$.
The only other possible pole for $\Tgt$ is at $\tau=0$ since the fundamental
region ${\cal F}_p=H/\Gp$ for $\Gp$ has only these two cusp points \Gunning.
{}From (7), $\Tgt$ has a pole at $\tau=0$ if and only if $E_g^0< 0$.
Thus $\Tgt$ is a hauptmodul for $\Gp$ if and only if $E_g^0\geq 0$.
Also from (7), $T_g(W_p(\tau))=\MMtextorbsqr{1}{g}(p\tau)$, so
that $\Tgt$ is a hauptmodul for $\Gp+$ if and only if $E_g^0=-1/p$ and $N_g=1$.

For classes of type $p+$,  $\Tgt= \MMtextorbsqr{1}{g}(p\tau)$
is a series in $q$ with non-negative coefficients since the RHS of (7) is the
$\Vg$ \pf . For classes of type $p-$,  $\Tgt$ has
coefficients of mixed sign. In general, all classes of $M$ can
be divided into two such types i.e. classes with Thompson series invariant
(or not invariant) under a Fricke involution $W_N:\tau\rightarrow -1/N\tau$
which are called Fricke (or non-Fricke)
classes. There are a total of 121 Fricke classes all of which have non-negative
coefficient Thompson series and 51 non-Fricke classes with mixed sign
coefficients for similar reasons to the prime ordered classes described.
This division of the classes of $M$ will be important below.

\subsect {The FLM Uniqueness Conjecture.}
 FLM have conjectured that $\MM$ is characterised (up to isomorphism) as
follows \FLM: {\it $\MM$ is the unique meromorphic conformal field theory with
modular invariant partition function $J(\tau)$ and central charge 24}.
This is analogous to the uniqueness property of the
Leech lattice as being the only even self-dual lattice in 24 dimensions without
roots.

Let us now consider orbifold models based on other \autos $a$ of the
untwisted Leech lattice theory $\VL$ lifted from \autos $\a\in {\rm Co_0}$
\refs{\Tuitetwo,\DongMason}. $\a$ will be chosen so that
each model contains no massless operators, has a meromorphic OPA and is
modular invariant with partition function $J(\tau)$ and hence,
should reproduce $\MM$. Each $\a\in {\rm Co_0}$ can be parameterised
as follows
$$
\eqalignno{
\det{x-\a}=&\prod_{k\vert n}(x^k-1)^{a_k}&(8a)\cr
\sum_{k\vert n}a_k=&0&(8b)\cr}
$$
with $\sum_{k\vert n} ka_k=24$ where $k\vert n$ denotes that $k$ divides $n$,
 the order of $\a$ and $\{a_k\}$ are integers. (8b) is imposed to ensure the
absence of fixed points for $\a$ so that no massless \ops in $\VL$ survive
the $\Pa$ projection. For $n=p$ prime, we have
$a_p=-a_1=2d$ where $(p-1)2d=24$.

Since $a$ is an OPA \auto for $\VL$, the $a$ invariant
subspace $\Pa\VL$ also forms a closed meromorphic OPA. The
partition function $\Tr{\Pa\VL}{q^{L_0}}$ is not modular invariant, as before,
necessitating the introduction of  sectors $\Va$ twisted by $a$.
Thus under  $\tau\rightarrow -1/\tau$
$$
\eqalign{
\orbsqr{a}{1}={1\over \eta_\a(\tau)}\rightarrow
\orbsqr{1}{a}=D_a^{1/2}
\prod_{k\vert n}\eta(\tau/k)^{-a_k}
=D_a^{1/2}q^{E_0^a}(1+O(q^{1/n}))
}
\eqno (9)
$$
with $\eta_\a(\tau)=\prod_k \eta(k\tau)^{a_k}$ and
$D_a={\rm det}(1-\a)$  where $D_a^{1/2}$ and
$E_0^a=-{1\over 24}\sum_k {a_k\over  k}$ are the degeneracy and
energy of the $a$ twisted vacuum. Under $\tau\rightarrow\tau+n$,
the $a$ twisted partition function is invariant up to a phase
$\exp(2\pi i n E_0^a)$. For modular consistency of the orbifold partition
function we must have $n E_0^a=0{\ \rm mod\ }1$ i.e. there is no global phase
anomaly \Vafa. In addition, if $E_0^a >0$, then the $a$ twisted sector does not
reintroduce massless states. We therefore restrict ourselves to  $\a\in
\rm Co_0$ obeying \Tuitetwo
$$
\eqalignno{
\sum_{k\vert n}a_k=&0 &(10a)\cr
E_0^a>&0&(10b)\cr
nE_0^a=&0 {\rm \  mod\ }1&(10c)\cr
}
$$
There are a total of 38 classes of $\rm Co_0$ \Kondo\ that obey
these constraints \Tuitetwo. If we relax condition (10c) then a further 13
classes of $\rm Co_0$  obey only (10a-b) \refs{\Tuitethree,\Unique}.
Each of these 13 classes is characterised
by some $h\not=1$ where $h\vert 24$ with $h\vert k$ for all $a_k\not =0$.
In all 51 cases the parameters $\{a_k\}$ obey  $a_k=-a_{nh/k}$ and so
$E_0^a=1/nh$ which violates (10c) for $h\not=1$.
$\textorbsqr{a}{1}$ is invariant up to phases of order $h$ under $\Gn$ and is a
hauptmodul for $\Ga$ with $\GN\subseteq\Ga\subset {\cal N}(N)$, $N=nh$,
where $\Ga$ is one of the genus zero modular groups considered by
Conway and Norton. Furthermore, since $E_0^a>0$, $\textorbsqr{a}{1}$ cannot
be Fricke invariant and hence these 51 hauptmoduls
are the 51 non-Fricke Monster group hauptmoduls. Thus there
is a correspondence between  51 classes $\{ \a\}$ of $\rm Co_0$  and the 51
non-Fricke classes of $M$.  We will explicitly identify an element $g_n\in M$
of each such class below.

$\Va$  with the \pfu $\textorbsqr{1}{a}$ of (9) has a standard construction
 \refs{\Twist}. Likewise, $\Vak$
twisted sectors must be introduced for modular invariance and OPA closure.
Then the following intertwining non-meromorphic OPA should hold (schematically)
$$
\eqalign{
\psi_{a^j}\psi_{a^k}\sim \psi_{a^{j+k}}
}
\eqno (11)
$$
with $\psi_{a^k}(z)\in \Vak$. Apart from the original $Z_2$ case,
this OPA has only been rigorously constructed in the prime ordered cases
\DongMason. We will assume that it is true in general. We therefore enlarge
$\VL$ by the introduction of $\Vak$ to
$\Vp=\VL\oplus\Va\oplus...\oplus {\cal V}_{a^{n-1}}$ which forms a closed
non-meromorphic OPA. The  projection $\Vorb=\Pa\Vp$ then forms a
meromorphic OPA. (10c) is sufficient to guarantee the modular
invariance of the corresponding \pf. (10b) can be also shown
to be sufficient to ensure no massless \ops appear in $\Pa\Vak$
\refs{\Tuitetwo, \Unique}. Thus, for the 38 \autos obeying (10a-c),
the \pfu is modular invariant and is given by $Z_{\rm orb}(\tau)=J(\tau)$.
Therefore $\Vorb\equiv\MM$ according to the FLM uniqueness
conjecture.  Let us now consider some evidence to support this.

Let $\Morb$ be the \auto group of the OPA for $\Vorb$ where we expect
$M=\Morb$. We define $i_a\in \Morb$ of order $n$
(which generalises the involution $i$ in the original FLM
construction) under which all the \ops of $\Vak$ are eigenstates with
eigenvalue $e^{2\pi i k/n}$.  $i_a$ is also an \auto of  $\Vp$
and is \lq dual\rq\ to the \auto $a$ where $\Pa\Vp=\Vorb$ and
${\cal P}_{i_a}\Vp=\VL$. Furthermore, we may reorbifold $\Vorb$ with
respect to $i_a$ to reproduce $\VL$ as before \Unique\
$$
\eqalign{
\matrix{
 \Vp  \cr
{\buildrel {\cal P}_{i_a}\ \ \over\swarrow} \qquad {\buildrel\ \
\Pa\over\searrow}  \cr
 \VL\quad
         \matrix{{\buildrel a\over \longrightarrow}  \cr
                {\buildrel i_a\over \longleftarrow}}
\quad\Vorb\cr}
}
\eqno (12)
$$
Thus if $\Vorb\equiv\MM$, we can explicitly construct the twisted
sectors ${\cal V}_{i_a^k}$ assumed earlier for $i_a\in M$.
We may also compute the Thompson series
for $i_a\in \Morb$ by taking the trace over $\Vorb$ to obtain
$$
\eqalign{
\Torb{i_a}=\Tr{\Vorb}{i_a q^{L_0}}={1 \over \eta_\a(\tau)}-a_1
}
\eqno(13)
$$
which is the hauptmodul for the genus zero modular group $\Ga$ introduced
earlier \Tuitetwo. Thus each $i_a\in \Morb$ dual to $a$ has the same
Thompson series as a corresponding non-Fricke element of $M$ e.g. for $\a$
of prime order $p$, $\Torb{i_a}=[\eta(\tau)/\eta(p\tau)]^{2d}+2d=
T_{p-}(\tau)$. Note also, from (7), that ${\cal V}_{i_a}$ has vacuum energy
$E^{i_a}_0=0$ and degeneracy $-a_1>0$.
(13) may be generalised to the other 13 classes
violating (10c) where $\a^h$, of order $n'=n/h$, can
be  employed to construct an orbifold with partition function $J(\tau)$. Let
$g_n$ denote the lifting of $\a$ where $g_n^h=i_{a^h}$ is dual to $a^h$,
a lifting of $\a^h$ (for $h=1$, $g_n=i_a$). We may then compute the
Thompson series for $g_n$ as a trace over ${\cal V}_{\rm orb}^{a^h}$
to show that (13) again holds so that $g_n$ has the same
Thompson series as a non-Fricke element of $M$ \Unique.

We may also compute the centraliser $C(g_n\vert M_{\rm orb}^{a^h})=
\{g\in M_{\rm orb}^{a^h}\vert g_n^{-1}gg_n=g\}$. For the 38 classes with
$h=1$ this consists of \autos that do not mix the sectors $\Pa\Vak$.
For the other 13 \autos $g_n$, $C(g_n\vert M_{\rm orb}^{a^h})\subset
C(i_{a^h}\vert M_{\rm orb}^{a^h})$. In general, $c\in C(i_a\vert \Morb)$
 must commute with $a$ and therefore $c$ is
lifted from the \auto $\c\in G_n=C(\a\vert{\rm Co}_0)/<\a>$. One can then show
that \refs{\Tuitethree,\Unique}
$$
\eqalign{
C(g_n\vert M_{\rm orb}^{a^h})=\Lahat.G_n}
\eqno (14)
$$
where $\Lahat=n.L_\a$, an extension of $L_\a=\L/(1-\a)\L$ by a cyclic group
of order $n$. $\Lahat$ arises from the vacuum structure
of $\Va$ where $D_a=\vert L_\a\vert$. With $\Morb=M$, (14) generalises a
a well-known observation of Conway and Norton concerning the
5 prime classes where $C(p-\vert M)=p_+^{1+2d}.G_p$ and $i_a=p-$ \CN.
For the other 46 classes, there are 11 cases
for which (14) can be checked using the
available information about these centralisers \refs{\CN,\Wilson}. In general,
the order of these groups agrees with (14) in each case supporting the
very likely validity of the result.

Both (13) and (14) support the conjecture that $\Vorb\equiv \MM$. This
can only be proved by finding a generalised version of $\sigma$
in the FLM construction which mixes the untwisted and
twisted sectors \refs{\FLM,\DGMtri} i.e. there should exist some
permutation group $\Sigma_n$ which mixes
the sectors of $\Vorb$ where $C(g_n\vert M)$ and $\Sigma_n$ generate
$M$. In the prime cases $p\not =2$, $\Sigma_p$ has been recently constructed
and it has been rigorously shown that $\Morb=M$ for $p=3$ and almost so for
$p=5,7,13$ \DongMason.

\subsect{Monstrous Moonshine from the Uniqueness of $\MM$.}
Let us now assume that the FLM Uniqueness conjecture is correct. We can then
argue
that {\it Thompson series are hauptmoduls if and only if orbifolding $\MM$ with
respect to elements of $M$ reproduces $\MM$ or $\VL$}. Thus Monstrous
Moonshine is intimately linked to the uniqueness of  $\MM$.

{}From (12), orbifolding $\MM$ with respect to the 38 non-Fricke
elements $i_a$ dual to $a$ reproduces $\VL$. We may similarly consider the
orbifolding of $\MM$ with respect to the Fricke elements $\{f\}$ with $h=1$
which
lead to a modular invariant theory $\MMforb$ \refs{\Tuiteone,\Unique}, given
that the \ops ${\cal V}_{f^k}$ can be constructed. Assuming that the Thompson
series are hauptmoduls we find that $\MMforb\equiv\MM$ i.e. orbifolding $\MM$
with respect to a Fricke \auto reproduces $\MM$ again. Thus we have \Unique
$$
\eqalign{
 \VL \
\matrix{{\buildrel a\over \longrightarrow}  \cr
            {\buildrel i_a\over \longleftarrow}}\
\MM\   {\buildrel f\over \longleftrightarrow}\ \MM
}
\eqno (15)
$$
For example, consider $f$ an element of a prime class $p+$. Fricke invariance
implies $\MMtextorbsqr{1}{f^k}=T_f(\tau/p)=q^{-1/p}+0+O(q^{1/p})$
so that there is a \lq gap\rq\ in the spectrum of ${\cal V}_{f^k}$ and no
massless
\ops are reintroduced in orbifolding $\MM$. Thus the modular invariant \pfu for
$\MMforb$ is $J(\tau)$ and hence $\MMforb\equiv\MM$.  A similar argument can
be made in the general case \Unique.

The converse to the above also holds i.e.
assuming that (15) is true for all \autos of $M$ that define a modular
consistent theory, then the Thompson series are hauptmoduls.
To see this, firstly consider an orbifolding with respect to $i_a\in M$ which
reproduces $\VL$. $i_a$ must be dual to one of the 38 \autos obeying (10a-c)
and has non-Fricke invariant Thompson series (13) which is the hauptmodul for a
genus zero group. Similarly, as discussed above, the other non-Fricke \autos
can also be found with a corresponding genus zero Thompson series. For the
remaining  Fricke classes of $M$ we provide an argument for
$f$ an element of prime order. We wish to show that $\Vf$ has the correct
vacuum structure so that $\Tft$ is a hauptmodul for $\Gp+$.
In the orbifolding of $\MM$ with respect to $f$ which reproduces $\MM$, let
$i_f\in M$ be dual to $f$ with eigenvectors $\Vfk$ for eigenvalue
$e^{2\pi ik/n}$. Then it can be shown that $T_{i_f}(\tau)=\Tft$ so
that $i_f$ is in the same class
as $f$. Furthermore, the centralisers obey $C(f\vert M)\subseteq
C(i_f\vert M)$ with the necessary equality only when the $\Vf$ vacuum is unique
i.e. $N_f=1$. Since the twisted sector $\Vf$ does not reintroduce massless
operators, the vacuum energy obeys either (a) $E^0_f=-1/p$ or (b) $E^0_f>0$.
(a) is possible because the absence of massless \ops in $\MM$ allows
for a similar \lq gap\rq\ in the spectrum of $\Vg$. If (b) holds, then $\Tft$
has a unique simple pole at $q=0$ and must be a hauptmodul for $\Gp$ with
$(p-1)\vert 24$ and $\Tft=[\eta(\tau)/\eta(p\tau)]^{2d}+2d$. However, this is
impossible since then $E^0_f=0$ with $N_f=2d$ from (7). Thus (15) implies that
$\Vf$ has vacuum structure $E^0_f=-1/f$ with $N_f=1$ and hence,
as described before, $\Tft$ is a hauptmodul for the genus zero group $\Gp+$
and $f$ is of class $p+$. A similar argument can be given for the other Fricke
classes \Unique.

\listrefs
\end